# Questioning the impact of AI and interdisciplinarity in science: Lessons from COVID-19

**Diletta Abbonato[a], Stefano Bianchini[a], Floriana Gargiulo[b], and Tommaso Venturini[c]**

[a] BETA – University of Strasbourg, France
[b] GEMASS-CNRS, Paris, France
[c] MEDIALAB – Université de Genève, Swiss

**Abstract.** Artificial intelligence (AI) has emerged as one of the most promising technologies to support COVID-19 research, with interdisciplinary collaborations between medical professionals and AI specialists being actively encouraged since the early stages of the pandemic. Yet, our analysis of more than 10,000 papers at the intersection of COVID-19 and AI suggest that these collaborations have largely resulted in science of low visibility and impact. We show that scientific impact was not determined by the overall interdisciplinarity of author teams, but rather by the diversity of knowledge they actually harnessed in their research. Our results provide insights into the ways in which team and knowledge structure may influence the successful integration of new computational technologies in the sciences.

Interdisciplinarity has become the buzzword in science policy. And with very good reason. Disciplines have for decades – in some cases centuries – facilitated scientific progress by providing scholars with the scaffolding of a coherent paradigm and with the possibility of standing on the shoulders of their predecessors. However, disciplinary boundaries have often proved to be a stumbling block to development, as growing specialization makes it ever harder (though ever more necessary) for scientists to venture into unexplored territories and combine practical and intellectual tools originating from different traditions (Jones, 2009). These entrenched boundaries are especially problematic when we find ourselves facing unprecedented research challenges that require fresh thinking and unrestrained experimentation.

Just such a situation presented itself recently with the outbreak of the COVID-19 pandemic. The urgency and gravity of the situation prompted researchers in epidemiology and medical science not only to mobilize all the resources available within their disciplines, but to look beyond them for new ideas and external collaborations. And among them, the alliance with artificial intelligence (AI) emerged as one of the most promising (Fig. 1).

**Figure 1. COVID-19 publications with AI content**

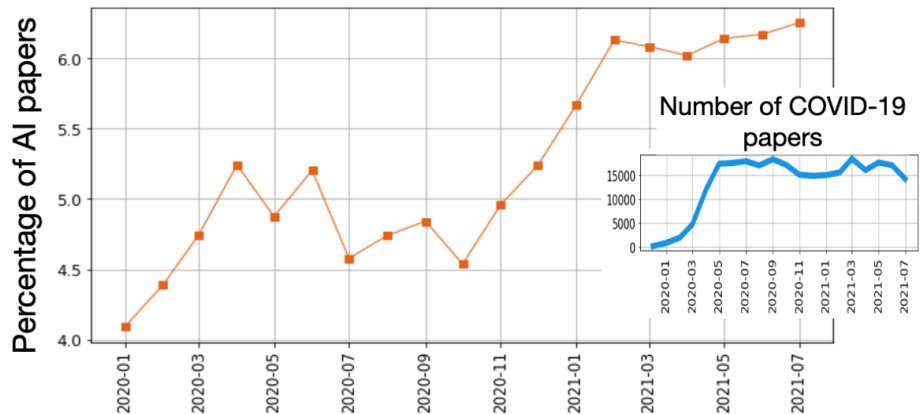

*Notes*: Fraction of COVID-19 papers containing AI. Inset: Total number of COVID-19 papers containing AI. After an initial period of exponential growth, scientific production related to the COVID-19 virus stabilized in May 2020. At the same time, AI research dedicated to COVID-19 virus remained relatively marginal until summer 2020 when it began to record constant linear growth, so that by July 2021 it accounted for nearly 7% of total COVID-19 scientific production. *Source*: Own elaboration on *CORD-19* data.

Although AI is nothing new, the field has recently been revived by the burgeoning power of computational technologies and the growing availability of data on social and natural phenomena. This has led to the development of new machine learning approaches, which are yielding remarkable results within and beyond data science (Cardon et al., 2018; Frank et al., 2019). The scientific enterprise is no exception to this trend. Some recent studies have shown that AI techniques are indeed changing the "way of doing science," from agenda setting and hypothesis formulation to experimentation, knowledge sharing, and public involvement, with a far from negligible impact on scientific discovery (Agrawal et al., 2018; Cockburn et al., 2018; Furman and Teodoridis, 2020; Bianchini et al., 2022).

The coronavirus pandemic hit at the peak of this cycle of AI hype and, unsurprisingly, many scholars quickly embraced ideas of adopting AI techniques to tackle the many challenges presented by COVID-19 (DeGrave et al., 2021; Khan et al., 2021; Roberts et al., 2021). Opportunities for collaborative funding have emerged globally to bring various scientific communities together, and researchers from different backgrounds have come together to try to harness the potential of AI in COVID-19 research (Ahuja et al., 2020; Luengo-Oroz et al., 2020). Yet, while some collaborations have made substantial contributions to the fight against the pandemic, others never got beyond the blueprint stage. What can explain these contrasting outcomes?

**Interdisciplinary research: Pros and Cons.** Previous research shows that (large) interdisciplinary teams produce more cited research and high-impact papers (Wuchty et al., 2007; Fortunato et al., 2018), and that diversity – not only epistemic, but also institutional and ethnic – is beneficial for producing novel, valuable ideas (Taylor and Greve, 2006). Teams comprising researchers with different backgrounds, methodological approaches, and experience have access to a broader pool of knowledge, which allows them to produce more creative outputs than those produced by traditional, non-collaborative science (Stephan, 2012; Uzzi et al., 2013; Gargiulo et al., 2022). Collaborative projects also serve to boost visibility by exposing scientific findings to a wider and more diverse readership (Leahey, 2016). How does this relate to COVID-19 research? Well, it suggests that collaborations between AI experts and clinicians may have mainly resulted in successful research outcomes, as domain specialists could provide their "on-the-ground" knowledge to identify promising areas for investigation related to the virus and related problems, while technology experts could apply the latest methods. A winning strategy, in short.

However, team diversity can also increase the chances of failure in collaborative research. Teams that are too large and heterogeneous often suffer from lower consensus-building, cognitive diversity, higher coordination costs, and emotional conflict. Thus, as diversity increases, it becomes more difficult to convert specialized expertise into scientific outputs (Lee et al., 2015). Some studies show that a team's ability to perform well depends more on how the team interacts than on the characteristics of its members (Woolley et al., 2010), and that most successful collaborations seem to be achieved through efforts that, while interdisciplinary, span relatively close fields (Yegros et al., 2015). Therefore, it is possible that conflicts could have arisen in collaborations between AI and COVID-19 experts due to differences in their areas of expertise, and this could have resulted in less impactful and visible scientific outcomes compared to teams consisting of only AI or clinical specialists.

The ultimate impact of interdisciplinarity remains an empirical question, one that we address in this paper. Here, based on a sizeable corpus of scientific publications at the intersection of COVID-19 and AI (~10,000 papers retrieved from the COVID-19 Open Research Dataset, *CORD-19* – version 2021-08-09 – and supplemented by other metadata from *Altmetric*, *OpenAlex*, and *Semantic Scholar*), we study which forms of interdisciplinarity served as the main drivers of scientific impact.

In the remainder, we first describe the metrics of interdisciplinarity that we devised for our study, and then link these metrics to three indicators of scientific "success", namely the number of citations, online visibility, and outreach to other disciplines.

**Measuring interdisciplinarity.** Each document, *i*, in our data is characterized by a set of authors ($\mathcal{A}_i$), a set of references and citations ($\mathcal{R}_i, \mathcal{C}_i$), a set of AI keywords, if any, ($\mathcal{W}_i$), the journal in which the paper is published ($\mathcal{J}_i$), and an altmetric score ($\mathcal{M}_i$). At the same time, for each author, *a*, present in our corpus we identified his/her list of papers ($\mathcal{P}_a$) and the list of his/her three most recent papers ($\mathcal{P}_a^3$).

Based on the co-occurrence of journals, in all the papers' reference lists, we employed a measure based on pairwise mutual information, to identify a distance matrix, D, among all the journals appearing in the dataset (if two journals are cited together several times their distance is considered small).

With this information, we defined two types of interdisciplinary metrics to evaluate the disciplinary positioning of each AI-COVID-19 paper: the first is related to team composition (measuring the difference in the scientific disciplinary background of the authors contributing to the paper); the second is related to the knowledge mobilized in the paper, in terms of reference heterogeneity.

For each dimension (team and knowledge), we develop a further distinction between metrics concerning AI ($\mu_{AI}^{team}$ and $\mu_{AI}^{kn}$) and those concerning their more general interdisciplinary nature ($\mu_{GEN}^{team}$ and $\mu_{GEN}^{kn}$), providing us with four different metrics:

- *AI team metric* is the fraction of previous AI publications for each author, averaged over the entire team:

$$\mu_{AI}^{team}(i) = \frac{1}{\#\mathcal{A}_i} \sum_{a \in \mathcal{A}_i} \frac{\#\{j \in \mathcal{P}_a \mid \mathcal{W}(j) \neq \{\}\}}{\#\mathcal{P}_a}$$

- *AI knowledge metric* is the fraction of cited references related to AI:

$$\mu_{AI}^{kn}(i) = \frac{\#\{j \in \mathcal{R}_i \mid \mathcal{W}(j) \neq \{\}\}}{\#\mathcal{R}_i}$$

- *General team metric* is the average disciplinary dispersion (in term of journal distances) of team authors:

$$\mu_{GEN}^{team}(i) = \frac{1}{\#\mathcal{A}_i} \sum_{a \in \mathcal{A}_i} \left( \frac{1}{3} \sum_{k \neq l \in \mathcal{P}_a^3} \mathbf{D}_{\mathcal{J}(k)\,\mathcal{J}(l)} \right)$$

- *General knowledge metric* is the average distance among all the journals cited in the references:

$$\mu_{GEN}^{kn}(i) = \frac{1}{\#(\mathcal{R}_i \times \mathcal{R}_i)} \sum_{(u,v) \in (\mathcal{R}_i \times \mathcal{R}_i)} \mathbf{D}_{\mathcal{J}(u)\,\mathcal{J}(v)}$$

To be clear, the first two metrics, $\mu_{AI}^{team}$ and $\mu_{AI}^{kn}$, measure the share of AI in the author teams and knowledge mobilized by the publications in our corpus, respectively. The remaining two, $\mu_{GEN}^{team}$ and $\mu_{GEN}^{kn}$, measure levels of general interdisciplinarity in the teams and knowledge, respectively.

For all the papers, we define three different indicators of 'success', namely: the *number of citations*, $\mathcal{N}(i)$, the *altmetric score*, $\mathcal{M}(i)$, and the *interdisciplinary spread* – i.e., how a paper is cited in a diverse set of disciplines – defined as:

$$\mathcal{I}(i) = \frac{1}{\#(\mathcal{C}_i \times \mathcal{C}_i)} \sum_{(u,v) \in (\mathcal{C}_i \times \mathcal{C}_i)} \mathbf{D}_{\mathcal{J}(u)\,\mathcal{J}(v)}$$

By topic modelling on abstracts of the papers in our corpus, we obtained five distinct application areas, which we label as follows: (i) *Societal Issues* (including epidemiology and infodemics); (ii) *Medical Imaging*; (iii) *Diagnosis and Prognosis*; (iv) *Treatments and Vaccines*; and (v) *Public Health*, with the most frequent uses of AI being found in medical imaging, followed by public health and, to a lesser extent, societal issues (Fig. 2).

**Figure 2. AI application areas for COVID-19 research**

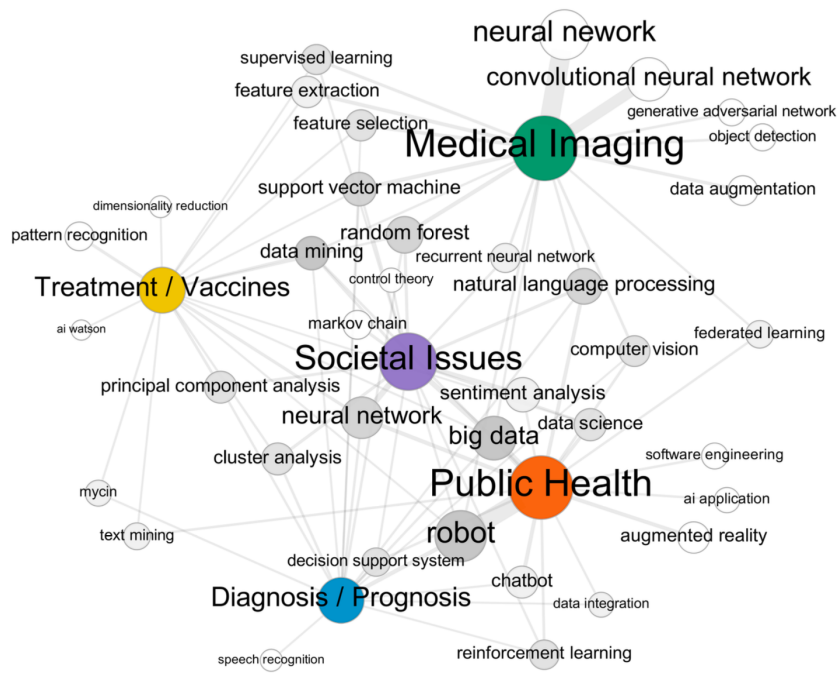

*Notes*: Co-occurrence of AI keywords (gray nodes) and COVID-19 topics (colored nodes). Edges are weighted by the number of articles that use each keyword in each topic. Nodes are sized according to their popularity (number of articles). Keywords are colored according to their degree, the number of topics in which each keyword is used (white keywords are specific to one topic, dark gray keywords are used in multiple topics).

A closer reading of the terms characterizing each topic suggests that AI has found a multitude of applications (Bullock et al., 2020; Naudé, 2020; Yang et al., 2020; Piccialli et al., 2021). In the case of societal issues, AI seems to have been used mainly for predicting the spread of disease over time and space, modeling public policy interventions (e.g., social distancing) and risk assessment, and fighting misinformation and disinformation on social media. In the case of medical imaging, what we essentially see is the deployment of deep learning models (e.g., CNN) to detect signs of COVID-19 from X-ray images and computed tomography (CT) scans. Another area of application, particularly of machine learning and deep learning, is the identification of possible treatments and vaccines, as well as the re-purposing of existing drugs. Finally, AI appears to support the management of the public health system, with robotics providing assistance in the delivery of healthcare tasks.

Each application area may have required specific skills and know-how from researchers with diverse backgrounds and experience with AI technology, and not least, the (re)combination of different types of knowledge. Unsurprisingly, our corpus reveals a high level of general interdisciplinarity both in the teams and in the knowledge mobilized by the publications across all research topics – with a slightly higher knowledge heterogeneity in medical imaging and diagnosis and prognosis (Fig. 3 top). In the case of AI, we observe very different scenarios at the topic level. Indeed, the share of teams with more AI experts is markedly higher in medical imaging and public health research, whereas teams working on vaccines, treatments, and prognosis seem to rely very little on AI knowledge (Fig. 3 bottom).

**Figure 3. Interdisciplinarity metrics in the different axes of COVID-19 research**

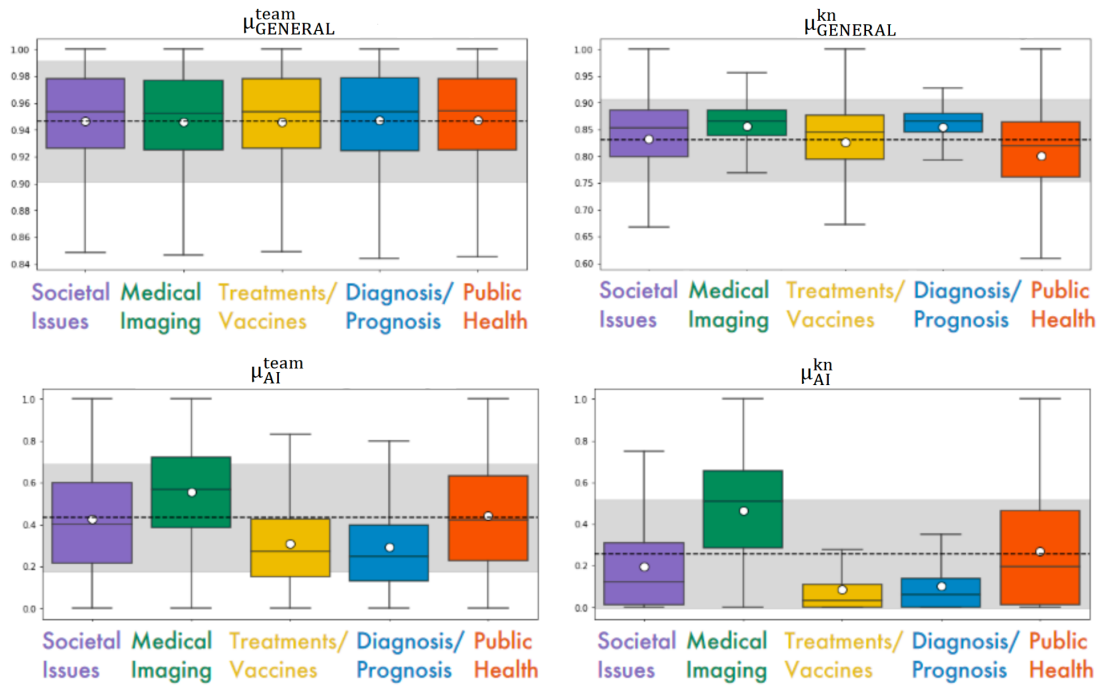

*Notes*: general (top) and AI-related interdisciplinarity (bottom). The dotted line and shaded area represent the mean and standard deviation, respectively.

**What determines 'success'.** We modeled the various impact measures – i.e., the number of citations received by the publication, the Altmetric attention score, and the interdisciplinary spread – as a function of four interdisciplinarity metrics discussed earlier – $\mu_{AI}^{team}$, $\mu_{AI}^{kn}$, $\mu_{GEN}^{team}$, and $\mu_{GEN}^{kn}$ – and a set of control variables, namely: *AI Collaborator* (=1 if the team includes at least one AI researcher), *Top AI Collaborator* (=1 if the team includes an AI researcher with past number of citations in the top 10° percentile of the citation distribution); *Academic Age* (average academic age of team members, in logs); *Past Impact* (average H-Index of team members based on past publications, in logs); *Nb. Countries* (number of participating countries within a team, in logs); and *Nb. References* (number of cited references, in logs). We also included a complete set of fixed effects for the month of publication and the dominant topic.

**Table 1. Determinants of 'success'**

|  | *Nb. Citations* (1) | *Attention Score* (2) | *Interd. Spread* (3) |
|---|---|---|---|
| $\mu_{AI}^{team}$ | -0.346*** (0.069) | -0.583*** (0.060) | -0.097*** (0.021) |
| $\mu_{AI}^{kn}$ | 0.083 (0.063) | -0.521*** (0.055) | -0.017 (0.019) |
| $\mu_{GEN}^{team}$ | 0.482* (0.271) | 0.259 (0.238) | -0.060 (0.080) |
| $\mu_{GEN}^{kn}$ | 2.165*** (0.176) | 2.162*** (0.150) | 0.299*** (0.054) |
| AI Collaborator | 0.055 (0.162) | -0.399*** (0.141) | -0.034 (0.050) |
| Top AI Collaborator | 0.298*** (0.076) | -0.070 (0.068) | 0.041* (0.023) |
| Past Impact (log) | 0.148*** (0.007) | 0.143*** (0.006) | 0.012*** (0.002) |
| Academic Age (log) | -0.265*** (0.025) | -0.231*** (0.022) | -0.033*** (0.007) |
| Nb. Countries (log) | 1.068*** (0.032) | 0.423*** (0.028) | 0.114*** (0.010) |
| Nb. References (log) | 0.385*** (0.015) | 0.068*** (0.013) | 0.063*** (0.004) |
| Observations | 12,180 | 12,180 | 8,734 |
| Log Likelihood | -38,868 |  |  |
| AIK | 77,816 |  |  |
| Adjusted R2 |  | 0.183 | 0.068 |
| F Statistic |  | 71.16*** | 17.33*** |

*Notes*: The statistical model for evaluating the relationship of different interdisciplinary metrics on three indicators of 'success': the number of citations received by the publication (Column 1), the Altmetric attention score (Column 2) and the interdisciplinary spread (Column 3). Coefficient estimates of time and topic fixed effects have been omitted from the table.

As shown in Table 1, the most notable result to emerge from our model is that collaborations with researchers experienced in AI (*AI Collaborator*) do not have a significant impact, and those involving a high share of researchers with established track records of AI publications ($\mu_{AI}^{team}$) receive, on average, fewer citations, have less online visibility, and struggle to reach distant disciplines. Only those teams that include a top AI researcher (*Top AI Collaborator*) present a positive impact on citations received by their publication, albeit that this impact is not strong. Similarly, the ratio of AI-related references ($\mu_{AI}^{kn}$) has a null or negative impact on the Altmetric attention score. All in all, research interdisciplinarity limited to AI does not seem to have any influence on the impact of COVID-19 publications, and when it does, this influence is negative.

What appears to ensure the impact of a publication is, above all else, the interdisciplinarity of the knowledge mobilized via its references ($\mu_{GEN}^{kn}$), that is the actual epistemological diversity of the research conducted by a team. This variable has a very strong positive effect on the number of citations and on the online visibility of a publication and this effect is consistently higher than that of more classic features, such as past impact or the number of affiliated countries. The overall diversity of team members ($\mu_{GEN}^{team}$) has only a marginal positive effect on the number of citations.

**Discussion.** The COVID-19 pandemic sparked a global research effort to address this unprecedented event. The scientific system responded promptly to the early stages of the virus and the international scientific community called upon its diverse expertise to assess the clinical and pathogenic characteristics of the disease and to formulate therapeutic strategies. Policymakers were also quick to seek advice from ethicists, sociologists, and economists on how best to deal with the crisis (Fry et al., 2020; Chahrour et al., 2020). Against this backdrop, AI applications represented a promising approach to face many of the challenges posed by the pandemic.

A number of studies focused on the AI-COVID-19 nexus have identified various barriers that may well have impeded the disciplines support of COVID-19 research. They include poor data quality and flow, as well as deficient global standards and database interoperability (e.g., genetic sequences, protein structures, medical imagery and epidemiological data); the inability of algorithms to work without sufficient knowledge of the domain; overly exacting computational, architectural, and infrastructural requirements; and the legal and ethical opacity associated with privacy and intellectual property (Bullock et al., 2020; Luengo et al., 2020; Naudé, 2020; Khan et al., 2021; Piccialli et al., 2021).

Here, we have analyzed the role played by different forms of interdisciplinarity, both at the team level and in the research conducted, and their repercussions on various measures of scientific impact. Our research was, in part, motivated by the fact that policy initiatives around the world have emerged – and continue to emerge – aimed at bringing the AI community and the healthcare system closer together. However, we have no direct evidence of the effectiveness of these initiatives. Our study provides an unequivocal takeaway message for academic decision-makers: collaborations involving AI researchers did not result in more impactful science, quite the contrary. What generates high-impact outcomes is not "on paper" interdisciplinarity engendered by team diversity, but rather the epistemological diversity hardwired into a paper. So how can team members best mobilize and blend ideas, tools, and knowledge from their scientific fields? We believe that further needs to comprehend the optimal team composition, conditions, and attributes for successful integration of novel computational technologies into scientific practices.

**Acknowledgment.** This work was supported by the La Mission pour les Initiatives Transverses et Interdisciplinaires (MITI) of the Centre National de la Recherche Scientifique (CNRS). This work was also supported by the European Union – Horizon 2020 Program under the scheme "INFRAIA-01-2018-2019 – Integrating Activities for Advanced Communities", Grant Agreement n.871042, "SoBigData++: European Integrated Infrastructure for Social Mining and Big Data Analytics" (http://www.sobigdata.eu). Stefano Bianchini received financial support through the SEED project – Grant agreement ANR-22-CE26-0013.

# References

Agrawal, A., McHale, J., & Oettl, A. (2018). Finding needles in haystacks: Artificial intelligence and recombinant growth. In *The economics of artificial intelligence: An agenda* (pp. 149-174). University of Chicago Press. https://doi.org/10.3386/w24541

Ahuja, A. S., Reddy, V. P., & Marques, O. (2020). Artificial intelligence and COVID-19: A multidisciplinary approach. *Integrative Medicine Research*, *9*(3). https://doi.org/10.1016/j.imr.2020.100434

Bianchini, S., Müller, M., & Pelletier, P. (2022). Artificial Intelligence in science: An emerging general method of invention. *Research Policy*, *51*(10), 104604. https://doi.org/10.1016/j.respol.2022.104604

Bullock, J., Luccioni, A., Pham, K. H., Lam, C. S. N., & Luengo-Oroz, M. (2020). Mapping the landscape of artificial intelligence applications against COVID-19. *Journal of Artificial Intelligence Research*, *69*, 807-845. https://doi.org/10.1613/jair.1.12162

Cardon, D., Cointet, J. P., & Mazières, A. (2018). Neurons spike back. *Réseaux*, *211*(5), 173-220. https://doi.org/10.3917/res.211.0173

Chahrour, M., Assi, S., Bejjani, M., Nasrallah, A. A., Salhab, H., Fares, M., & Khachfe, H. H. (2020). A bibliometric analysis of COVID-19 research activity: A call for increased output. *Cureus*, *12*(3). 10.7759/cureus.7357

Cockburn, I. M., Henderson, R., & Stern, S. (2018). The impact of artificial intelligence on innovation *(No. w24449)*. *National Bureau of Economic Research*. https://www.nber.org/papers/w24449

DeGrave, A. J., Janizek, J. D., & Lee, S. I. (2021). AI for radiographic COVID-19 detection selects shortcuts over signal. *Nature Machine Intelligence*, *3*(7), 610-619. https://doi.org/10.1038/s42256-021-00338-7

Fortunato, S., Bergstrom, C. T., Börner, K., Evans, J. A., Helbing, D., Milojević, S., ... & Barabási, A. L. (2018). Science of science. *Science*, *359*(6379). 10.1126/science.aao0185

Frank, M. R., Wang, D., Cebrian, M., & Rahwan, I. (2019). The evolution of citation graphs in artificial intelligence research. *Nature Machine Intelligence*, *1*(2), 79-85. https://doi.org/10.1038/s42256-019-0024-5

Fry, C. V., Cai, X., Zhang, Y., & Wagner, C. S. (2020). Consolidation in a crisis: Patterns of international collaboration in early COVID-19 research. *PloS One*, *15*(7). https://doi.org/10.1371/journal.pone.0236307

Furman, J. L., & Teodoridis, F. (2020). Automation, research technology, and researchers' trajectories: Evidence from computer science and electrical engineering. *Organization Science*, 31(2), 330–354. https://doi.org/10.1287/orsc.2019.1308

Gargiulo, F., Castaldo, M., Venturini, T., & Frasca, P. (2022). Distribution of labor, productivity and innovation in collaborative science. *Applied Network Science*, *7*(1), 1-15. https://doi.org/10.1007/s41109-022-00456-0

Jones, B. F. (2009). The burden of knowledge and the "death of the renaissance man": Is innovation getting harder?. *The Review of Economic Studies*, 76(1), 283–317. https://www.jstor.org/stable/20185091

Khan, M., Mehran, M. T., Haq, Z. U., Ullah, Z., Naqvi, S. R., Ihsan, M., & Abbass, H. (2021). Applications of artificial intelligence in COVID-19 pandemic: A comprehensive review. *Expert Systems with Applications*, *185*, 115695. https://doi.org/10.1016/j.eswa.2021.115695

Leahey, E. (2016). From sole investigator to team scientist: Trends in the practice and study of research collaboration. *Annual Review of Sociology*, *42*, 81-100. https://doi.org/10.1146/annurev-soc-081715-074219

Lee, Y. N., Walsh, J. P., & Wang, J. (2015). Creativity in scientific teams: Unpacking novelty and impact. *Research Policy*, *44*(3), 684-697. https://doi.org/10.1016/j.respol.2014.10.007

Luengo-Oroz, M., Hoffmann Pham, K., Bullock, J., Kirkpatrick, R., Luccioni, A., Rubel, S., ... & Mariano, B. (2020). Artificial intelligence cooperation to support the global response to COVID-19. *Nature Machine Intelligence*, *2*(6), 295-297. https://doi.org/10.1038/s42256-020-0184-3

Naudé, W. (2020). Artificial intelligence vs COVID-19: Limitations, constraints and pitfalls. *AI & Society*, *35*(3), 761-765. https://doi.org/10.1007/s00146-020-00978-0

Piccialli, F., Di Cola, V. S., Giampaolo, F., & Cuomo, S. (2021). The role of artificial intelligence in fighting the COVID-19 pandemic. *Information Systems Frontiers*, *23*(6), 1467-1497. https://doi.org/10.1007/s10796-021-10131-x

Roberts, M., Driggs, D., Thorpe, M., Gilbey, J., Yeung, M., Ursprung, S., ... & Schönlieb, C. B. (2021). Common pitfalls and recommendations for using machine learning to detect and prognosticate for COVID-19 using chest radiographs and CT scans. *Nature Machine Intelligence*, *3*(3), 199-217. https://doi.org/10.1038/s42256-021-00307-0

Stephan, P. (2015). *How economics shapes science*. Harvard University Press.

Taylor, A., & Greve, H. R. (2006). Superman or the fantastic four? Knowledge combination and experience in innovative teams. *Academy of Management Journal*, *49*(4), 723-740. https://www.jstor.org/stable/20159795

Uzzi, B., Mukherjee, S., Stringer, M., & Jones, B. (2013). Atypical combinations and scientific impact. *Science*, 342(6157), 468–472. DOI: 10.1126/science.1240474

Yang, G. Z., J. Nelson, B., Murphy, R. R., Choset, H., Christensen, H., H. Collins, S., ... & McNutt, M. (2020). Combating COVID-19—The role of robotics in managing public health and infectious diseases. *Science Robotics*, *5*(40). DOI: 10.1126/scirobotics.abb5589

Yegros-Yegros, A., Rafols, I., & D'este, P. (2015). Does interdisciplinary research lead to higher citation impact? The different effect of proximal and distal interdisciplinarity. *PloS One*, *10*(8). https://doi.org/10.1371/journal.pone.0135095

Wang, D., & Barabási, A. L. (2021). *The science of science*. Cambridge University Press.

Wang, L. L., Lo, K., Chandrasekhar, Y., Reas, R., Yang, J., Eide, D., ... & Kohlmeier, S. (2020). CORD-19: The Covid-19 open research dataset. *arXiv preprint arXiv:2004.10706v4*. https://arxiv.org/abs/2004.10706

Woolley, A. W., Chabris, C. F., Pentland, A., Hashmi, N., & Malone, T. W. (2010). Evidence for a collective intelligence factor in the performance of human groups. *Science*, *330*(6004), 686-688. DOI: 10.1126/science.1193147

Wuchty, S., Jones, B. F., & Uzzi, B. (2007). The increasing dominance of teams in production of knowledge. *Science*, *316*(5827), 1036-1039. DOI: 10.1126/science.1136099

# Supplementary information

**Data.** Our analysis combines data from four different databases – *CORD-19, Semantic Scholar, OpenAlex*, and *Altmetric* – and is based on the pre-processing protocol illustrated in Fig. 1A.

The COVID-19 Open Research Dataset (CORD-19) is a growing corpus of publications on COVID-19 and other coronavirus infections (Wang et al., 2020). It includes, in the period that we considered (from 01/12/2019 to 31/08/2021), around 600K documents from different sources, including WHO, PubMed central, bioRxiv and medRxiv.

Within this large corpus, we focused specifically on a subset of 26,887 publications that included, in the abstract or in the title, at least one keyword related to AI. Our list of around 300 AI keywords (Tab. 1A) was retrieved by merging the terms mentioned in the Wikipedia AI Glossary for AI with several other pages entitled as 'AI vocabulary' and 'AI glossary on the web'.

For each paper in this subset, we retrieved additional metadata from Semantic Scholar and OpenAlex. We discarded any documents with missing information and obtained a final corpus of 16,148 AI publications on COVID-19 (COVID-19+AI dataset). For each of these papers, we then retrieved all their references (c. 1 million papers) and all papers citing them (c. 200K papers), as well as the metadata associated with all these papers.

Semantic Scholar metadata included the DOI, which we used for retrieving the 'attention score' for each paper in the COVID-19+AI dataset from the website Altmetric.com. This provides a measure of online activity for scholarly content (e.g., mentions on the news, in blogs, and on Twitter; article page-views and downloads; GitHub repository watchers). We used the author identifier in OpenAlex to retrieve the previous publications of all 87,552 authors present in our corpus (around 150K papers) and the institutions to which they are affiliated.

**Figure 1A. Data preparation pipeline**

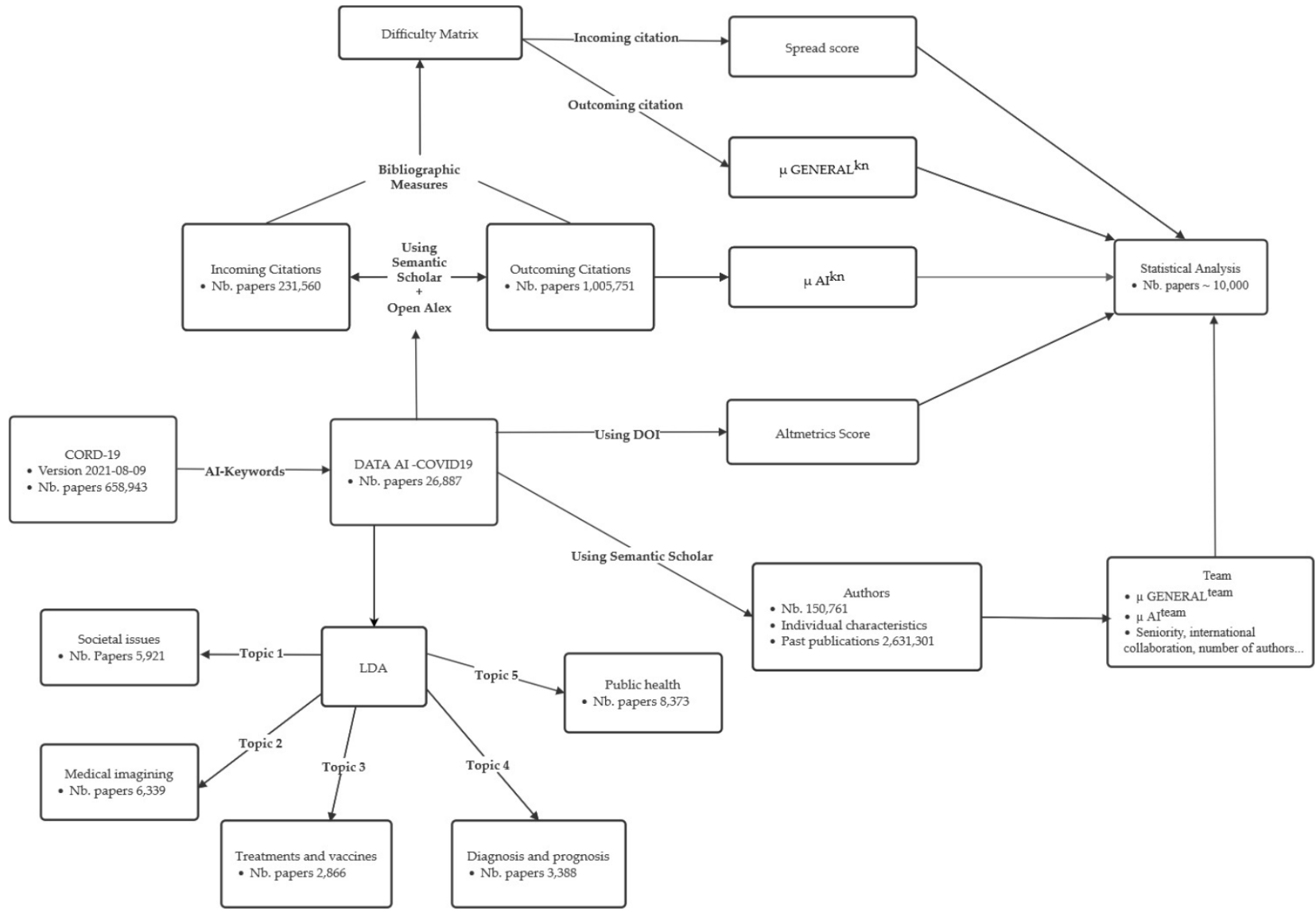

# Table 1A. AI terms for document retrieval

| | | | | |
|---|---|---|---|---|
| abductive logic programming | boolean satisfiability problem | developmental robotics | kl one | ontology learning |
| abductive reasoning | brain technology | dialogue system | knowledge acquisition | open mind common sense |
| abstract data type | branching factor | dimensionality reduction | knowledge engineering | openai |
| action language | brute-force search | discrete system | knowledge extraction | opencog |
| action model learning | capsule neural network | distributed artificial intelligence | knowledge interchange format | partial order reduction |
| action selection | case based reasoning | dynamic epistemic logic | knowledge representation andreasoning | partially observable markov decision process |
| activation function | chatbot | eager learning | knowledge-based system | particle swarm optimization |
| adaptive algorithm | cloud robotics | ebert test | lazy learning | pathfinding |
| adaptiveneuro fuzzy inference system | cluster analysis | echo state network | lisp | pattern recognition |
| admissible heuristic | cobweb | embodied agent | logic programming | predicate logic |
| adversarial neural | cognitive architecture | embodied cognitive science | long short term memory | predictive analytics |
| affective computing | cognitive computing | ensemble averaging | machine learning | principal component analysis |
| agent architecture | cognitive science | error driven learning | machine listening | principle of rationality |
| ai accelerator | combinatorial optimization | ethics of artificial intelligence | machine perception | probabilistic programming |
| ai application | committee machine | evolutionary algorithm | machine translation | prolog |
| ai applications | commonsense knowledge | evolutionary computation | machine vision | propositional calculus |
| ai complete | commonsense reasoning | evolving classification function | markov chain | qualification problem |
| aiml | computational chemistry | existential risk from artificial general intelligence | markov decision process | quantum computing |
| alphago | computational complexity theory | expert system | mathematical optimization | query language |
| ambient intelligence | computational creativity | fast and frugal trees | mechanism design | radial basis function network |
| answer set programming | computational cybernetics | feature extraction | mechatronics | random forest |
| anytime algorithm | computational humor | feature learning | meta learning | reasoning system |
| application programming interface | computational intelligence | feature selection | metabolic network reconstruction and simulation | recurrent neural |
| approximate string matching | computational learning theory | federated learning | metaheuristic | recurrent neural network |
| approximation error | computational linguistics | first order logic | model checking | region connection calculus |
| argumentation framework | computational mathematics | forward chaining | modus ponens | reinforcement learning |
| artificial general intelligence | computational neuroscience | friendly artificial intelligence | modus tollens | reservoir computing |
| artificial immune system | computational number theory | fuzzy control system | monte carlo tree search | resource description framework |
| artificial intelligence | computational problem | fuzzy logic | multi agent system | restricted boltzmann machine |
| artificial neural network | computational statistics | fuzzy rule | multi swarm optimization | rete algorithm |
| association for the advancement of artificial intelligence | computer automated design | fuzzy set | mycin | robot |
| asymptotic computational complexity | computer vision | general game playing | naive bayes classifier | robotics |
| attributional calculus | concept drift | generative adversarial network | naive semantics | rule-based system |
| augmented reality | connectionism | genetic algorithm | name binding | satisfiability |
| automata theory | consistent heuristic | genetic operator | named entity recognition | search algorithm |
| automated planning and scheduling | constrained conditional model | glowworm swarm optimization | named graph | self-management |
| automated reasoning | constraint logic programming | graph database | natural language | semantic analysis |
| autonomic computing | constraint programming | graph theory | natural language generation | semantic network |
| autonomous car | constructed language | graph traversal | natural language processing | semantic query sensor fusion |
| autonomous robot | control theory | halting problem | natural language programming | semantic reasoner |
| backpropagation | convolutional | hyper heuristic | network motif | semantic search |
| backpropagation through time | convolutional neural | ieee computational intelligence society | neural machine translation | semi supervised learning |
| backward chaining | convolutional neural network | image detection | neural network | sentiment analysis |
| bag of words model | darkforest | image recognition | neural networking | separation logic |
| bag of words model in computer vision | dartmouth workshop | incremental learning | neural networks | similarity learning |
| batch normalization | data augmentation | inference engine | neural turing machine | situation calculus |
| bayesian programming | data fusion | information integration | neuro fuzzy | speech recognition |
| bees algorithm | data integration | intelligence amplification | neuromorphic engineering | statistical learning |
| behavior informatics | data mining | intelligence explosion | nlp | supervised learning |
| behavior tree | data science | intelligent agent | nondeterministic algorithm | tensorflow |
| belief desire intention software model | datalog | intelligent control | nouvelle ai | text mining |
| bias-variance tradeoff | decision boundary | intelligent machine | np completeness | trajectory forecasting |
| big data | decision support system | intelligent personal assistant | np hardness | trasnfer learning |
| big o notation | deep learning | issue tree | object detection | unsupervised learning |
| binary tree | deepmind technologies | junction tree algorithm | occam's razor | |
| blackboard system | default logic | keras | offline learning | |
| boltzmann machine | description logic | kernel method | online machine learning | |

**Table 2A. Most recurrent bigrams in the corpus**

| Aggregate | | Societal Issues | | Medical Imagining | | Treatments and Vaccine | | Diagnosis and Prognosis | | Public Health | |
|---|---|---|---|---|---|---|---|---|---|---|---|
| public health | 2748 | social medium | 1600 | chest xray | 1973 | molecular docking | 588 | mental health | 653 | health care | 885 |
| social medium | 2283 | public health | 1412 | xray image | 1762 | main protease | 435 | systematic metanalysis | 565 | public health | 738 |
| chest xray | 2011 | confirmed case | 841 | ct image | 1228 | immune response | 388 | risk factor | 539 | mental health | 522 |
| xray image | 1782 | social distancing | 795 | ct scan | 1117 | spike protein | 324 | intensive car | 419 | social medium | 433 |
| mental health | 1614 | infectious disease | 601 | chest ct | 1022 | molecular dynamic | 305 | controlled trial | 405 | internet thing | 416 |
| health care | 1607 | united states | 563 | transfer learning | 948 | signaling pathway | 286 | clinical trial | 396 | contact tracing | 400 |
| result show | 1576 | number case | 403 | computed tomography | 903 | drug discovery | 274 | logistic regression | 337 | social distancing | 398 |
| learning algorithm | 1446 | mental health | 390 | learning model | 842 | amino acid | 271 | health care | 335 | digital technology | 371 |
| learning model | 1424 | using learning | 375 | learning algorithm | 697 | drug repurposing | 271 | mechanical ventilation | 332 | digital health | 337 |
| using learning | 1411 | case death | 359 | experimental result | 675 | clinical trial | 261 | viral infection | 254 | language processing | 303 |
| social distancing | 1363 | air quality | 345 | learning method | 652 | healthcare system | 312 | cross-sectional study | 305 | health system | 292 |
| infectious disease | 1334 | march 2020 | 307 | tomography ct | 639 | binding affinity | 243 | care unit | 302 | virtual reality | 271 |
| ct image | 1285 | health organization | 304 | using learning | 596 | gene expression | 236 | significant difference | 296 | face mask | 267 |
| chest ct | 1257 | world health | 301 | chest x-rays | 522 | innate immune | 226 | included study | 296 | clinical trial | 262 |
| ct scan | 1236 | learning model | 289 | learning approach | 505 | virtual screening | 220 | confidence interval | 295 | learning algorithm | 256 |
| learning method | 1139 | reproduction number | 282 | learning technique | 465 | chinese medicine | 214 | study conducted | 295 | medical student | 252 |
| confirmed case | 1120 | learning algorithm | 276 | publicly available | 463 | recognition receptor | 213 | analysis performed | 285 | higher education | 251 |
| computed tomography | 1082 | language processing | 258 | sensitivity specificity | 449 | antiviral drug | 202 | public health | 280 | infectious disease | 243 |
| learning approach | 1061 | learning approach | 254 | chain reaction | 408 | immune system | 192 | statistically significant | 280 | distance learning | 227 |
| transfer learning | 1048 | learning technique | 251 | cxr image | 395 | component analysis | 191 | material method | 280 | supply chain | 226 |

*Notes*: This table reports the 20 most recurrent non-AI (see Table 1A) bigrams in publication corpus, aggregate and by dominant topic

**The statistical model.** The empirical analysis tests the determinants of the various impact measures, our three dependent variables: the number of citations received by the publication, the Altmetric attention score and the interdisciplinary spread (measuring how publications are cited in a diverse set of disciplines). The number of citations, $\mathcal{N}$, is a count variable and was modelled using a negative binomial regression. The continuous variables – attention score $\mathcal{M}$ and interdisciplinarity spread $\mathcal{I}$ – were modeled using ordinary least square regressions.

For each paper, we included other factors in the models that could influence its impact and visibility: namely *AI Collaborator* (=1 if the team includes at least an AI researcher); *Top AI Collaborator* (=1 if the team includes an AI researcher with past number of citations in the top 10° percentile of the citation distribution); *Academic Age* (average academic age of team members, in logs); *Past Impact* (average H-Index of team members based on past publications, in logs); *Nb. Countries* (number of participating countries within a team, in logs); and *Nb. References* (number of cited references, in logs). We also included a complete set of dummies for the month of publication and the dominant topic.

**Some geographical trends.** While COVID-19 article production is geographically distributed according to the general patterns of scientific productivity observed in previous studies (Fry et al., 2020; Wang et al., 2021) (with the United States, the United Kingdom and China leading the way, followed by Western European countries and India), the use of AI in COVID-19 research presents a different distribution (Fig. 2A). The countries of Asia and the Middle East – China and India, in particular – appear as leaders of AI-based COVID-19 research, while the USA and Western European countries lag someway behind.

**Table 2A. Geographical distribution of publication activity AI-COVID-19**

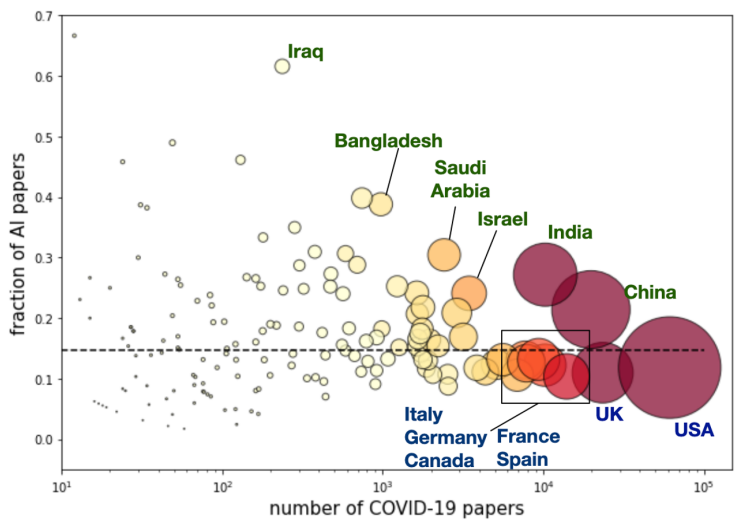

*Notes*: Plot A: Fraction of COVID-19 papers containing AI. Inset: Total number of COVID-19 papers containing AI. Plot B: Fraction of COVID-19 papers containing AI by country. Nodes are sized and colored according to the total number of COVID-19-AI papers. The dotted line represents the sample average.